\documentclass[twocolumn,prl,aps,showpacs,superscriptaddress]{revtex4-1}
\usepackage{graphicx}
\usepackage{color}
\usepackage{dcolumn}
\usepackage{amsmath}

\newlength{\figwidth} 
 \newlength{\figwidthb} %
\newcommand{\SIO}{Sr$_3$Ir$_2$O$_7$ }
\newcommand{\SIOns}{Sr$_3$Ir$_2$O$_7$}
\newcommand{\SIOs}{Sr$_2$IrO$_4$ } 
\newcommand{\SIOsns}{Sr$_2$IrO$_4$} 
\newcommand{\jeff}{$J_{\textrm{eff}}$}

\begin{document}

\title{Dimensionality Driven Spin-Flop Transition in Layered Iridates}

\author{J. W. Kim} 
\author{Y. Choi} 
\author{Jungho Kim}
\affiliation{Advanced Photon Source, Argonne National Laboratory, Argonne, Illinois 60439, USA}
\author{J. F. Mitchell}
\affiliation{Materials Science Division, Argonne National Laboratory, Argonne, IL 60439, USA}
\author{G. Jackeli}
\affiliation{Max Planck Institute for Solid State Research, Heisenbergstra\ss e 1, D-70569 Stuttgart, Germany}
\author{M. Daghofer}
\author{J. van den Brink}
\affiliation{Institute for Theoretical Solid Sate Physics, IFW
  Dresden, Helmholtzstra\ss e 20, 01069 Dresden, Germany}
\author{G. Khaliullin}
\affiliation{Max Planck Institute for Solid State Research, Heisenbergstra\ss e 1, D-70569 Stuttgart, Germany}
\author{B. J. Kim}
\affiliation{Materials Science Division, Argonne National Laboratory, Argonne, IL 60439, USA}

\date{\today}

\begin{abstract}
Using resonant x-ray diffraction, we observe  
an easy $c$-axis collinear antiferromagnetic structure for the bilayer \SIOns,
a significant contrast to the single layer \SIOs with in-plane canted moments. 
Based on a microscopic model Hamiltonian, we show that 
the observed spin-flop transition as a function of number of IrO$_2$ layers is
due to strong competition among intra- and inter-layer bond-directional
pseudo-dipolar interactions of the spin-orbit entangled \jeff=1/2 moments. With this we unravel the origin of anisotropic exchange interactions in a Mott insulator in the strong spin-orbit coupling regime, which holds the key to the various types of unconventional magnetism proposed in 5$d$ transition metal oxides.
\end{abstract}


\maketitle

Despite the long history of research on magnetism in insulating oxides,
magnetism in 5$d$ transition-metal oxides (TMO) with strong spin-orbit
coupling (SOC) is only now beginning to be explored. Since the recent discovery of the SOC-driven Mott insulator with \jeff=1/2 states in \SIOsns~\cite{bjkim08,bjkim09},
a wide array of theoretical proposals have been put forward for novel types of
quantum magnetism and topological phases of
matter~\cite{jackeli09,BalentsNPhys10,chaloupka10,WanPRB11,wang11,HyartPRB12,Witczak12}. The 
magnetism in the strong SOC limit has two fundamentally novel
aspects: (i) orbitals of different symmetries are admixed by SOC and thus the
magnetic exchange interactions are multidirectional, which is evident in 
particular from the ``cubic'' shape of the \jeff=1/2  Kramers doublet
wavefunction relevant for tetravalent iridates~\cite{jackeli09,bjkim08,bjkim09}; (ii) the
quantum phase inherent in the \jeff=1/2 states can strongly suppress the
isotropic Heisenberg coupling via a destructive interference among multiple
superexchange paths, and lead to large anisotropic exchange couplings,
of the form of pseudo-dipolar (PD) and Dzyaloshinsky-Moriya (DM)
interactions~\cite{jackeli09}. This provides a mechanism for frustrated 
magnetic interactions that are predicted to lead to unconventional
magnetism, such as the Kitaev model with spin liquid ground
state~\cite{jackeli09,chaloupka10,SinghPRL12}. By contrast, magnetic
interactions in the weak SOC limit are predominantly of isotropic Heisenberg
type weakly perturbed by the anisotropic couplings.  

The central theoretical premise underlying various iridates is that the
Kramers pair of \jeff=1/2 states is the correct starting point. Strictly speaking, however, the exact \jeff=1/2 states are realized
only in cubic symmetry {\it and} in the large Coulomb correlation
limit. Although it has been shown that in \SIOsns, having tetragonal symmetry
at the Ir site, the ground state wave function is indeed close to the \jeff=1/2
state~\cite{bjkim09}, it is not {\it a priori} obvious that this should also be
the case for other iridates with symmetries lower than cubic. Further, the
\jeff=1/2 states are also perturbed by the hopping term, the effect of which
should be more pronounced in iridates with small charge gap such as the
\SIOns~\cite{MoonPRL09}, a bilayer variant of the single layer
\SIOsns. Experimentally, a clear signature of the unique features of the
interactions inherent to the \jeff=1/2 moments, {\it e.g.} strong PD
couplings, has yet to be seen, especially in
(Na,Li)$_2$IrO$_3$~\cite{HillPRB11,ChoiPRL12,YePreprint12}, the candidate 
material for realization of the Kitaev model. 

In this Letter, we report a direct manifestation of the strong PD interactions
in \SIOns, which result from the \jeff=1/2 states that are robust despite the
proximity of \SIO to the metal-insulator transition (MIT) boundary. 
Using resonant x-ray diffraction (RXD), we find in \SIO a $G$-type antiferromagnetic (AF) structure
with $c$-axis collinear moments, in contrast to $ab$-plane canted
AF found for \SIOsns. The observed spin-flop transition as a function of
number of IrO$_2$ layers does not accompany an orbital reconstruction, which
shows that the strong inter-layer PD couplings, supported by the 
three-dimensional (3D) shape of the \jeff=1/2 wavefunction, 
are indeed responsible for the spin-flop transition. Employing the microscopic
model Hamiltonian of Ref.~\cite{jackeli09}, we show that in wide -- and 
realistic -- parameter ranges, the same microscopic 
parameters describing the \jeff=1/2 electronic states lead
to easy-$ab$-plane moments for the single layer \SIOs and $c$-axis collinear
moments for \SIOns. This implies that the transition occurs only as a function
of dimensionality, which is a consequence of the robustness of \jeff=1/2
states (albeit perturbed to some extent) against strong quasi-3D hopping
amplitudes. 

\begin{figure}[t]
\hspace*{-0.1cm}\vspace*{-0.2cm}\centerline{\includegraphics[width=0.8\columnwidth,angle=0]{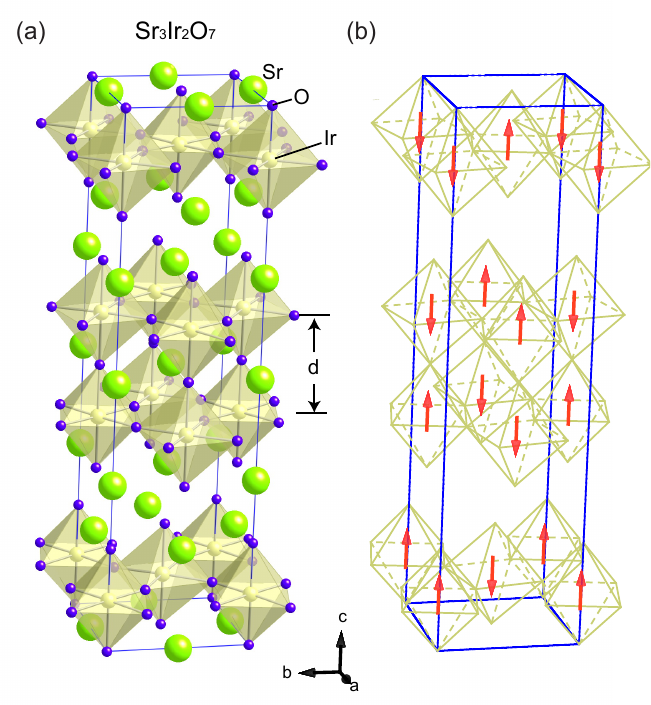}}  

\caption{(a) Crystal structure of \SIO as reported in Ref.~\cite{cao02}. Every
  neighboring IrO$_6$ octahedra are rotated in opposite sense about the
  $c$-axis by $\simeq 12^\circ$. (b) Magnetic order has a $c$-axis collinear
  G-type antiferromagnetic structure. The up and down magnetic moments
  correlate with counter-clockwise and clockwise rotations of the IrO$_6$
  octahedra, respectively.}\label{fig:fig1} 
\end{figure}

Experiments were carried out at the 4-IDD and 6-ID beamlines at the
Advanced Photon Source, with incident photon energy tuned to Ir
L$_{2,3}$ edges. A horizontal scattering geometry was used with
$\pi$-polarized incident beam. The polarization of the scattered x-rays
was analyzed with pyrolytic graphite (0 0 8) and (0 0 10) reflections
for L$_3$ and L$_2$ edges, respectively. A single crystal was 
mounted on a closed-cycle cryostat and data were
collected at a temperature of about 5 K. No indications of change in
the magnetic structure were found in the measurements repeated at 120
K and 250 K. X-ray absorption spectra were recorded simultaneously in
partial fluorescence mode using an energy-dispersive detector. 

Figure 1 shows the magnetic structure solved in the present study
along with the underlying crystal structure. \SIO was
first reported to adopt the space group
$I4$/$mmm$~\cite{Subramanian94}, but was later assigned to
$Bbcb$ based on single crystal diffraction and transmission
electron microscopy~\cite{cao02,Matsuhata04,structure}. In this orthorhombic
structure, all neighboring octahedra are rotated in an opposite sense
about the $c$ axis, breaking inversion symmetries with respect to the
shared oxygen ions and thereby allowing DM interactions. 

The $c$-axis collinear AF structure [Fig. 1(b)] is unambiguously solved from analysis of data presented in
Fig.~2 and Fig.~3.  Figure 2(a) shows magnetic Bragg peaks scanned over a wide range of
$l$, with ($h$,$k$) fixed at (1,0) and (0,1). The crystallographyically forbidden $h+k= odd$
reflections imply AF ordering within an IrO$_2$ plane, and the
observed large intensity modulation along the $l$ direction reflects the
bilayer magnetic structure factor. The magnetic peaks were refined
at each $l$, and the corresponding intensities obtained from
integrating rocking curves are plotted in Fig.~2(b). The intensity
modulation has a periodicity set by the ratio between the lattice
parameter $c$ and the bilayer distance $d$ (see Fig. 1),
i.e. $c/d\approx5.13$, and agrees well with the profile expected for
AF ordering between two neighboring IrO$_2$ planes. Thus, it follows
that all nearest-neighbor pairs are AF ordered. Fig.~2(c) shows
the temperature dependence of the intensity of (0 1 19) reflection,
which disappears above $\approx$285 K and correlates with the reported
anomalies in the magnetization and the resistivity data~\cite{cao02}, implying that
these anomalies are associated with the onset of long range AF
ordering. 

%
%

\begin{figure}[t]
\hspace*{-0.2cm}\vspace*{-0.1cm}\centerline{\includegraphics[width=0.9\columnwidth,angle=0]{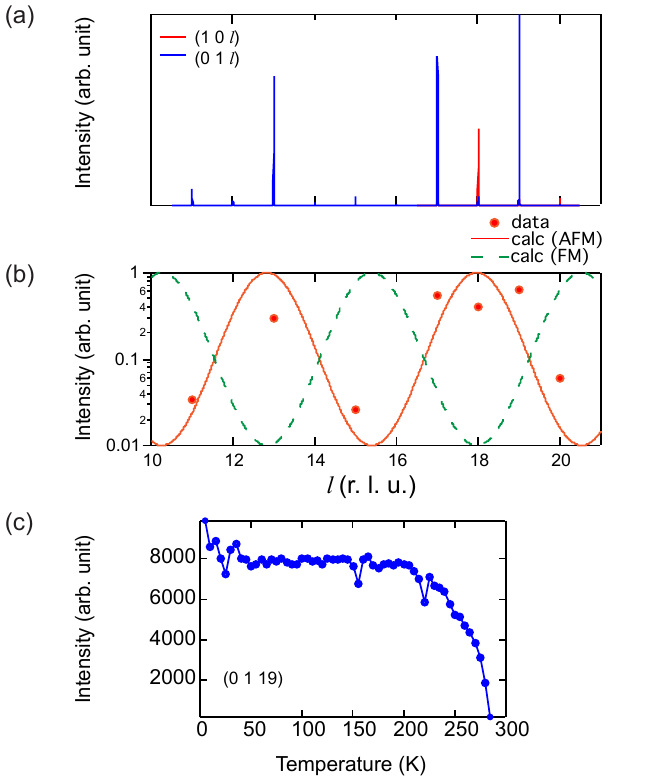}}

\caption{(a) $l$ scan measured in $\pi-\sigma$ polarization channel showing
  magnetic Bragg peaks. (b) Integrated intensities at each peak obtained from
  rocking curves (red dots). Red solid (green dashed) line is bilayer
  structural factor expected for antiferromagnetic (ferromagnetic) alignment
  of two adjacent IrO$_2$ planes in a bilayer expressed by $\cos^2\frac{2\pi
    d}{c}$ ($\sin^2\frac{2\pi d}{c}$). (c) Temperature dependence of (0 1 19)
  peak.}\label{fig:fig2} 
\end{figure}

%
%

\begin{figure}[t]
\hspace*{-0.2cm}\vspace*{-0.1cm}\centerline{\includegraphics[width=1\columnwidth,angle=0]{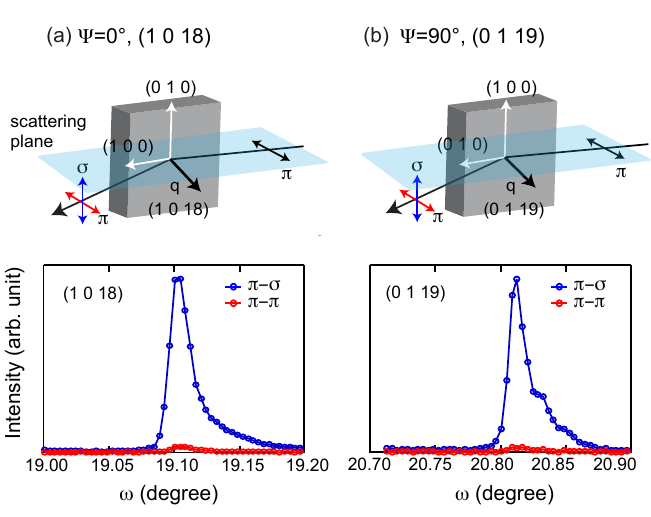}} 

\caption{Rocking curves measured in two polarization channels for (a) (1 0 18)
  reflection and (b) (0 1 19) reflections. The azimuth angle, defined with
  respect to the reference vector (1 0 0), was set at 0$^\circ$ for (1 0 18)
  reflection so that the scattering plane is defined by (1 0 0) and (1 0
  18). Likewise, for (0 1 19) reflection azimuth angle was set at 90$^\circ$
  so that the scattering plane is defined by (0 1 0) and (0 1 19).  
}\label{fig:fig3}  
\end{figure}

To determine the orientation of the magnetic moment, we performed polarization
analysis on two magnetic Bragg peaks, as shown in Fig.~3.  The (1 0 18)
reflection was recorded at the azimuthal angle $\Psi$=0$^\circ$ defined such
that it is zero when the reference vector (1 0 0) is in the scattering
plane. The data show that (1 0 18) reflection appears only in the $\pi-\sigma$
channel, demonstrating that the component of the magnetic moment contributing to
this reflection is confined to the scattering plane defined by (1 0 0) and (1
0 18) vectors. This implies the easy axis is in the $ac$ plane. Rotating
$\Psi$ by 90$^\circ$, now (0 1 0) and (0 1 19) vectors are contained in the
scattering plane. In this geometry, the (0 1 19) reflection also appears only
in the $\pi-\sigma$ channel, from which follows that the easy axis also lies in the
$bc$ plane. Taking these two data together, it is unambiguously determined
that the magnetic Bragg peaks of (1 0 $l$) and (0 1 $l$) are associated with
the $c$-axis component of the magnetic moment. With the provision of $\vec k$=0 ordering, the magnetic structure is uniquely
solved as shown in Fig. 1(b)~\cite{noteMagneticGroup}. 

%
%

\begin{figure}[t]
\vspace*{0.1cm}\centerline{\includegraphics[width=0.95\columnwidth,angle=0]{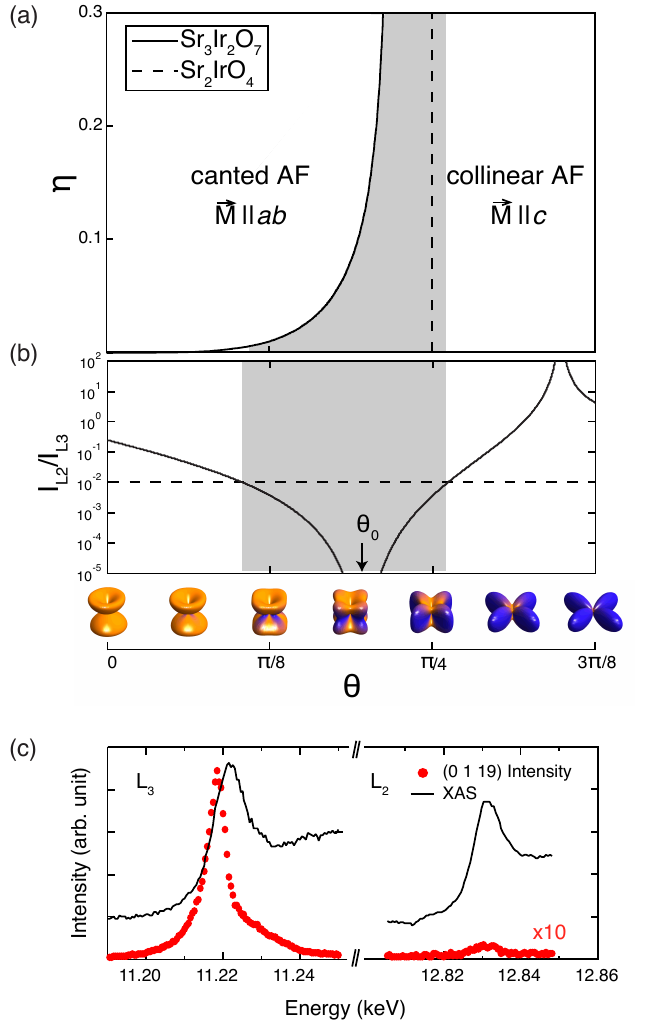}}
\caption{(a) The ground state phase diagram of the Hamiltonian (\ref{H}) in
  terms of $\eta=J_H/U$ and the tetragonal  
distortion parameter $\theta$ [the values of $\theta$ smaller (larger) than
$\theta_0$ correspond to compressed (elongated) octahedra]. The solid (dashed)
line marks the spin-flop transition in bilayer (single-layer) system. The
shaded area indicates the parameter space for \SIO constrained by experimental
observations (see text).  
(b) The ratio of intensities at L$_2$ and L$_3$
calculated as a function of $\theta$. The experimental ratio of at most ~1$\%$
provides the lower and upper bounds for $\theta$. Spin-orbital density map is
shown for some values of $\theta$ with spin up (down) represented by orange
(blue) color.  
(c) Energy scan of (0 1 19) reflection scanned around Ir L$_3$ and L$_2$
resonance. Red dots and black lines indicate scattering intensity and x-ray
absorption spectra, respectively. } 
\label{fig:fig4} 
\end{figure}

Having determined the magnetic structure, we now analyze the origin of
distinct magnetic orderings in layerd iridates. In the case of single 
layer \SIOsns, it has been shown~\cite{jackeli09} that DM couplings can be
gauged away by a proper rotation of quantization axes, and 
the magnetic anisotropy is solely decided by the bond-directional PD
interactions whose sign and hence the moment direction is controlled by the
tetragonal distortion parameter $\theta$ alone (while $\eta=J_H/U$, the ratio
of Hund's exchange and the local Coulomb repulsion, scales the magnitude of
the PD terms and magnon gaps). In the bilayer \SIO case, however,
one may expect strong inter-layer couplings since the spin-orbit entangled 
wavefunction in iridates is spatially of 3D 
shape~\cite{jackeli09,bjkim08,bjkim09}. 
This suggests that magnetic states in iridates may strongly vary with
dimensionality as number of planes are increased, unlike the case of cuprates
with spin-only moments that reside on planar orbitals.        

The magnetic interactions for intra- and inter-layer bonds of neighboring
iridium ions can be expressed in the following common form (with differing
coupling constants for inter- and intra-layer bonds): 

\begin{eqnarray}
{\cal H}_{ij}=J_{ij}\vec S_i\cdot \vec S_{j}+\Gamma_{ij}S_{i}^{z}S_{j}^{z}
+{\vec D}_{ij}\cdot{\big [}\vec S_i\times\vec S_j{\big ]} ,
\label{H}
\end{eqnarray} 

\noindent
where the first term  stands for isotropic AF exchange $J_{ij}$, the second
term describes symmetric anisotropy $\Gamma_{ij}$ that includes PD terms
driven by Hund's exchange and those due to staggered rotations of
octahedra~\cite{jackeli09}. The latter also induce  DM interaction, with DM
vector ${\vec D}_{ij}$ parallel to $c$-axis on all bonds, while its direction
is staggered. For intra-layer bonds, the coupling constants are identical to
those for the single layer case, derived in Ref.~\cite{jackeli09} in terms of
$\eta$, $\theta$, and the octahedra rotation angle $\alpha$.   We have
extended the same derivation to inter-layer bonds (the expressions are
somewhat lengthy and are given in the {\it Supplementary Material}). The
Hamiltonian (\ref{H}) supports two types of ordered states: a canted AF
structure with moments in $ab$-plane and a collinear AF order with moments
along $c$-axis, with AF  inter-layer stacking in both cases. We obtained a
classical phase boundary between these phases marking a spin-flop transition
as a function of $\eta$ and $\theta$. The result is shown as a solid line in 
Fig.~\ref{fig:fig4}(a). The dashed line in the same figure shows the spin-flop
transition for the single-layer compound. It is evident that the AF order 
with $c$-axis moments has 
a wider stability window in the bilayer compound than in the single-layer one,
which is primarily due to the PD interaction acting on the $c$-axis bond. 
One may note that the same set of
parameters in the wide parameter space, bounded by the solid and dashed lines 
in Fig.~4(a), leads to in-plane moments for single layer \SIOs and $c$-axis
moments for bilayer \SIOns; {\it i.e.}, the spin-flop transition occurs
without an accompanying ``orbital'' reconstruction. 

To gain insights into the competing magnetic interactions, consider first 
the case in which there is no octahedra rotation ($\alpha$=0). Whereas the 
intra-layer anisotropic coupling $\Gamma$ can change sign as a function of $\theta$, 
favoring $c$-axis collinear (in-plane canted) structure when $\theta$ is 
larger (smaller) than $\theta_c \approx\pi/4$~\cite{jackeli09}, the 
inter-layer PD term $\Gamma_c$ always favors $c$-axis collinear structure 
irrespective of the value of $\theta$, so
that there is a competition between intra- and inter-layer PD terms when
$\theta$$<$$\theta_c$ which is resolved in \SIO in favor of the c-axis
collinear AF order. We find that these PD terms can be considerably large
compared to the case of $3d$ TMOs, e.g., cuprates~\cite{PetitPRB99}; 
for realistic values of $\eta$
and $\theta$, the PD couplings can be as large as 0.1$\sim$0.3~$J$.
Now, restoring finite $\alpha$ we find that emerging DM interactions further
stabilize this structure. Indeed, as seen in the phase diagram [Fig. 4(a)], 
even at $\eta=0$ (in which case there would be no magnetic anisotropy in the 
single-layer case~\cite{jackeli09}) the AF order with $c$-axis moments 
is favored due to DM terms. The reason is that (in contrast to the
case of \SIOsns) the DM interactions in bilayer \SIO cannot be simultaneously 
gauged away, because the intra- and inter-layer bonds prefer different
canting angles.   

The observed dimensionality driven spin-flop transition, which is rare
in TMOs, is a natural consequence of the electronic ground state close to
\jeff=1/2 states with multidirectional [see Fig.~4(b)] and strong anisotropic 
couplings. Indeed, it is seen that \SIO has a similar degree of deviation
from the exact \jeff=1/2 states (for which L$_2$ RXD intensity is zero) as in
the single layer \SIOsns~\cite{bjkim09}, as evidenced by the smallness of L$_2$
intensity ($I_{{\textrm L}_2}/I_{{\textrm L}_3}$$<$1$\%$) as shown in
Fig.~4(c).  By contrast, in TMOs with polarized orbitals, adding another
layer would generally not affect the magnetic structure unless accompanied
by an orbital transition; for example in cuprates,
the planar $x^2$-$y^2$ orbitals cannot mediate strong anisotropic inter-layer
couplings.  
We find robust \jeff=1/2 states in \SIOns, a system lying 
close to the borderline of MIT with strong hopping amplitudes and 
a dimensionality greater than two. The validity of \jeff=1/2 picture
has also been confirmed recently~\cite{Takagi11} in CaIrO$_3$, a
post-perovskite material with edge-sharing geometry relevant to the Kitaev 
model, pointing out that the \jeff=1/2 states may be more generally 
applicable beyond the Ruddelsden-Popper series.  

It remains to be clarified how the observed $c$-axis collinear structure can
be reconciled with the reported unusual magnetic effects, such as weak
ferromagnetism, diamagnetism, and magnetoresistivity observed at rather low
magnetic fields below 1 T applied in the in-plane direction~\cite{cao02}. A
possible scenario is that the moment may be canted off from the $c$-axis with
the in-plane component appearing at different propagation vector
$q$'s. 
An alternative possibility is that an additional order parameter is present
and is responsible for the above magnetic effects. More investigations are
needed to resolve these issues. 

In summary, we have revealed -- through the observation of spin-flop 
transition in layered iridates -- a direct manifestation of the PD
interactions that are expected for \jeff=1/2 states and are the essential 
component of the unique magnetism proposed in 5$d$ TMOs with strong SOC.
For bilayer iridate \SIOns, these interactions lead to the collinear AF ground
state with moments directed along the $c$-axis, in contrast to easy-plane
canted AF structure of \SIOsns.  
The strong dependence of magnetic structure on the number of IrO$_2$ planes reflects the spin-orbit entangled nature of wavefunctions, which are
spatially of 3D shape and support strong inter-layer couplings,  
unlike the case of, {\it e.g.}, cuprates with planar orbitals. 
The resulting competition between intra- and inter- layer PD
(and also DM) interactions is tuned by the octahedral rotation and tetragonal
distortion, giving rise to the moment reorientation.  
Our experimental confirmation of robust \jeff=1/2 states in a system close to a MIT and their strongly non-Heisenberg interactions
strengthens the expectation for novel magnetism in correlated oxides with
strong SOC.     

\acknowledgments{The work in the
  Material Science Division and the use of the Advanced Photon Source at the
  Argonne National Laboratory was supported by the U.S. DOE under Contract
  No. DE-AC02-06CH11357. G.J. acknowledges support from GNSF/ST09-447,
  M.D. from the DFG (Emmy-Noether program).}  

\bibliography{sio327rxsnew}

\end{document}